\documentclass[pre,showpacs,preprint]{revtex4}

\usepackage{amsmath}

\usepackage{graphicx}
\usepackage{dcolumn}
\usepackage{bm}

\begin{document}

\title{Homogeneous hydrodynamics of a collisional model of confined granular gases }
\author{J. Javier Brey, P. Maynar, M.I. Garc\'{\i}a de Soria,  and V. Buz\'{o}n}
\affiliation{F\'{\i}sica Te\'{o}rica, Universidad de Sevilla,
Apartado de Correos 1065, E-41080, Sevilla, Spain}
\date{\today }

\begin{abstract}
The hydrodynamic equation governing the homogeneous time evolution of the temperature in a model of confined granular gas is studied by means of the Enskog equation. The existence of a normal solution of the kinetic equation is assumed as a condition for hydrodynamics. Dimensional analysis implies a scaling of the distribution function that is used to determine it in the first Sonine approximation, with a coefficient that evolves in time through its dependence on the temperature. The theoretical predictions are compared with numerical results obtained by the direct simulation Monte Carlo method, and a good agreement is found. The relevance of the normal homogeneous distribution function  to derive inhomogeneous hydrodynamic equations, for instance using the Champan-Enskog algorithm, is indicated.
\end{abstract}


\maketitle

\section{Introduction}
\label{s1}
Granular systems exhibit a rich phenomenology that is far from being well understood. When they are fluidized, many of the observed features have a strong similarity to what happens in ordinary, molecular fluids. Therefore, it is not surprising the great  deal of effort made to develop a hydrodynamic theory for granular gases, i.e. for granular systems under conditions in which  the particles move freely and independently between binary collisions \cite{Ha83,Go03}. Although the success of granular hydrodynamics is undeniable, there are still many open fundamental questions, whose answer can only be obtained by using kinetic theory and statistical mechanics methods \cite{ByP04,Du00,DByB08}.

Granular gases are inherent non-equilibrium systems due to inelasticity of collisions. There is no equilibrium state that can be used as a zeroth order or reference local state when developing a theory to describe the possible states of a system. Instead, there is a homogeneous state, with uniform hydrodynamic fields, which cools monotonically in time: the homogeneous cooling state (HCS) \cite{Ha83}. This essential feature has relevant implications in the structure of the macroscopic transport equations \cite{BDKyS98,DByB08,ByC01}.

A steady situation can be reached if energy is continuously supplied to a granular gas. In general, the price to be paid is that the system develops spatial inhomogeneities, in such a way that the energy dissipated in collisions is compensated by the energy flux associated to the hydrodynamic gradients. Nevertheless, it has been observed in experiments that it is possible to generate an {\em almost} uniform steady state of a granular gas by considering the horizontal dynamics of a vibrated system confined to a quasi-two-dimensional geometry \cite{OyU05,RPGRSCyM11}. The theoretical description of the dynamics of the  two-dimensional fluid generated in this way is a relevant physical question. Of course, the main issue is to describe how the energy is translated from the wall vibrating in the vertical direction to the horizontal degrees of freedom of the particles.  A first possibility is to consider an external noise force acting on each particle  \cite{vNyE98,PLMyV99,PTvNyE01}. Although it is true that this modeling leads to the existence of a uniform steady state, its possible relation with the state generated in the experiments with vibrated confined granular gases  has not been established.

Very recently, an alternative approach to the description of the two-dimensional dynamics of a confined granular gas has been proposed \cite{BRyS13,BGMyB13}.  The idea is to substitute the real three-dimensional collision rule by an effective two-dimensional one, trying to capture the mechanism for which the particles convert their kinetic energy associated to the vertical component of the velocity into kinetic energy in the horizontal plane. That makes a conceptually important difference with the stochastic models mentioned above. While in the latter the rate of injection of energy into the two-dimensional motion of the particles  is independent of their collision rate, in the model of  ref. \cite{BRyS13} it is precisely the collision rate what controls the interchange of energy between the vertical and horizontal degrees of freedom. For this reason, we refer to this model as to a collisional one. The main features of its homogeneous steady state were analyzed in ref. \cite{BGMyB13}. Here, the dynamics of the system towards the steady state, assuming that it remains always homogeneous, will be addressed.
The interest of this issue is twofold. Firstly, it will provide useful information to validate the model by comparing its predictions with molecular dynamics results of vibrated granular systems and even with experiments. Work is presently in progress along this line. Secondly, it permits to study hydrodynamics around non-equilibrium steady states in one of the simplest possible scenarios.

In ordinary fluids, the only translationally invariant hydrodynamic state is equilibrium. There is no possible hydrodynamic evolution of the temperature if the hydrodynamic fields are homogeneous. On the other hand, isolated  granular fluids have a time-dependent non-equilibrium homogeneous  state, the HCS, as mentioned above. The time evolution of the temperature  of a granular gas in this state obeys a macroscopic, hydrodynamic equation. When external energy is continuously supplied to a granular gas, a homogeneous steady state can be possible, depending on the way in which the energy is injected. This is what happens in the confined vibrated systems being modeled here. In this case, the existence of an homogeneous  hydrodynamic regime, characterized by a closed equation for the evolution of the temperature towards its steady value, can be expected on the appropriate time scale.  In the context of kinetic theory, hydrodynamics is guaranteed if the time evolution of the system is described by a {\em normal} velocity distribution, having the property that all its time dependence occurs through the granular temperature. This implies some scaling property of the distribution function. The particular form of the scaling depends on the number of parameters defining the system (or the model) and their nature. The case of a granular  gas driven by an external stochastic  thermostat has already been studied and the existence of homogeneous hydrodynamics established \ \cite{GMyT12,GMyT13}. In this work,  a similar analysis will be presented for the collisional model of confined granular gases.

It is worth to stress the relevance of this homogeneous time-dependent hydrodynamic state. When elaborating a theory for hydrodynamics in a confined granular gas, the general zeroth order in the gradients reference state to be considered is not a local version of the stationary state eventually reached by the system, but a distribution based on the time dependent state characterizing the homogeneous hydrodynamics. Only when the system is considered to be close to the steady situation, linearization around this state will be allowed. For general hydrodynamic states, an expansion around a local time-dependent state has to be considered. The reason for this is that a steady state occurs when the rate of variation of the temperature vanishes. This gives a relationship between the temperature and some parameters of the system. As a consequence, the temperature is not arbitrary but fixed in the steady state, and there is not a unique way of identifying the dependence on the temperature of the distribution function of that state. Consequently, it makes no sense to use this state to define arbitrary hydrodynamic fields as the starting point of the Chapman-Enskog method. This is a quite general feature of steady non-equilibrium states, as discussed in a seminal paper by Lutsko \cite{Lu06}.

The plan of the paper is as follows. In the next section, the model is formulated in a concise way, and some results previously derived are summarized. Assuming the existence of an homogeneous normal solution of the pseudo-Liouville equation of the system, formally exact equations are derived for the evolution of the temperature and the one-particle distribution function. The later is the first equation of an infinity hierarchy, as usual in non-equilibrium statistical mechanics.  In order to close it and get a kinetic equation,  the Enskog approximation is used in Sec. \ref{s3}. Moreover, calculations are restricted to the first Sonine correction to a Gaussian distribution. Then, it is possible to obtain an expression for the temperature rate and also an equation for the coefficient $a_{2}$ in  front of  the first Sonine correction. The equation has a special solution such that all the other solutions tend to it, before reaching the steady value. This kind of behaviour corresponds to the existence of hydrodynamics, in the sense that  all the time dependence of the distribution function occurs through the temperature.  In Sec. \ref{s4}, the theoretical predictions are compared with simulation results obtained by means of the direct simulation Monte Carlo method, and a quite good agreement is found. Finally, Sec. \ref{s4} contains a short summary and discussion of the main results.

\section{Homogeneous dynamics }
\label{s2}

The system considered is composed of $N$ identical hard spheres ($d=3$) or disks ($d=2$) of mass $m$ and diameter $\sigma$ enclosed in a volume $V$.
Although the particular application of the model in mind is restricted to a two-dimensional dynamics, it will be developed for arbitrary dimension. The reason is that it can be done with little additional  effort, and it could be useful in the future for other potential applications.  The position and velocity coordinates of the particles will be denoted by $\{ {\bm r}_{i},{\bm v}_{i}; i=1, \ldots, N \}$. The dynamics consists of free streaming until a given pair of particles $i,j$ are at contact. At this moment, the velocities of the two particles change instantaneously according to the inelastic collision rule \cite{BRyS13,BGMyB13}
\begin{equation}
\label{2.1}
{\bm v}_{i} \rightarrow {\bm v}^{\prime}_{i} = {\bm v}_{i}- \frac{1+ \alpha}{2}  {\bm v}_{ij} \cdot \widehat{\bm \sigma}  \widehat{\bm \sigma} + \Delta \widehat{\bm \sigma},
\end{equation}
\begin{equation}
\label{2.2}
{\bm v}_{j} \rightarrow {\bm v}^{\prime}_{j} = {\bm v}_{j}+ \frac{1+ \alpha}{2}  {\bm v}_{ij} \cdot \widehat{\bm \sigma} \widehat{\bm \sigma} - \Delta \widehat{\bm \sigma}.
\end{equation}
Here ${\bm v}_{ij} \equiv {\bm v}_{i}-{\bm v}_{j}$ is the relative velocity prior collision, $ \widehat{\bm \sigma}$ is a unit vector directed from the center of particle $j$ to the center of particle $i$ through the point of contact, and $\Delta$ is some positive characteristic speed.  The coefficient of normal restitution $\alpha$ takes values in the interval $0 < \alpha \leq 1$. Total momentum of the pairs is conserved in collisions, but there is a change in kinetic energy given by
\begin{equation}
\label{2.3}
e^{\prime}_{ij}-e_{ij} = m\left[ \Delta^{2} - \alpha \Delta  {\bm v}_{ij} \cdot \widehat{\bm \sigma} - \frac{1-\alpha^{2}}{4}\,  ({\bm v}_{ij} \cdot \widehat{\bm \sigma} )^{2} \right].
\end{equation}
The state of the system at time $t$ is completely characterized by the positions and velocities of all particles at that time and it is represented by a point $\Gamma_{t} \equiv \left\{ {\bm r}_{1}(t), \ldots {\bm r}_{N}(t), {\bm v}_{1}(t), \ldots, {\bm v}_{N}(t) \right\}$ in the associated $2dN$-dimensional phase space. The dynamics of the system corresponds to a deterministic trajectory in this phase space. In the context of statistical mechanics, the system is described by means of the probability density $\rho (\Gamma, t)$. In ref. \cite{BGMyB13}, the pseudo-Liouville equation obeyed by this function was derived,
\begin{equation}
\label{2.4}
\frac{\partial}{\partial t}\, \rho (\Gamma, t) = \overline{L}_{+}(\Gamma) \rho (\Gamma,t),
\end{equation}
with the pseudo-Liouville operator $\overline{L}_{+}(\Gamma)$ given by
\begin{equation}
\label{2.5}
\overline{L}_{+}(\Gamma) \equiv - \sum_{i=1}^{N} {\bm v}_{i} \cdot \frac{\partial}{\partial {\bm r}_{i}}+ \sum_{1 \leq i < j \leq N} \overline{T}_{+}(i,j).
\end{equation}
Here,  $\overline{T}_{+}(i,j)$ is the binary collision operator
\begin{eqnarray}
\label{2.6}
\overline{T}_{+}(i,j) & \equiv &  \sigma^{d-1} \int d \widehat{\bm \sigma}\, \left[ \theta({\bm v}_{ij} \cdot \widehat{\bm \sigma} - 2 \Delta) ({\bm v}_{ij} \cdot \widehat{\bm \sigma}  - 2 \Delta) \delta ({\bm r}_{ij} -{\bm \sigma}) \alpha^{-2} b_{\bm \sigma}^{-1} (i,j) \right. \nonumber \\
& & - \left. \theta ( {\bm v}_{ij} \cdot \widehat{\bm \sigma} ){\bm v}_{ij} \cdot \widehat{\bm \sigma}  \delta ({\bm r}_{ij} + {\bm \sigma}) \right].
\end{eqnarray}
In the above expression, ${\bm \sigma} \equiv \sigma  \widehat{\bm \sigma}$, $\theta$ is the Heaviside step function, and the operator $b_{\bm \sigma}^{-1}(i,j)$  changes all the velocities ${\bm v}_{i}$ and ${\bm v}_{j}$ to its right into the pre-collisional values,
\begin{equation}
\label{2.7} b_{\bm \sigma}^{-1} (i,j) {\bm v}_{i} = {\bm v}_{i} - \frac{1+ \alpha}{2 \alpha}\,  {\bm v}_{ij} \cdot \widehat{\bm \sigma}  \widehat{\bm \sigma} + \frac{\Delta \widehat{\bm \sigma}}{\alpha},
\end{equation}
\begin{equation}
\label{2.8} b_{\bm \sigma}^{-1} (i,j){\bm v}_{j} = {\bm v}_{j} + \frac{1+ \alpha}{2 \alpha}\, {\bm v}_{ij} \cdot \widehat{\bm \sigma}  \widehat{\bm \sigma} - \frac{\Delta \widehat{\bm \sigma}}{\alpha}\, .
\end{equation}
From the Liouville equation, the Born-Bogoliubov-Green-Kirkwood-Ivon (BBGKY) hierarchy of equations for the reduced distribution functions is easily  derived \cite{BGMyB13}. The first equation of the hierarchy reads
\begin{equation}
\label{2.9}
\left( \frac{\partial}{\partial t} +{\bm v}_{1} \cdot \frac{\partial}{\partial {\bm r}_{1}} \right) f_{1}({\bm r}_{1},{\bm v}_{1},t) = \int d{\bm r}_{2} \int d{\bm v}_{2}\,  \overline{T}_{+}(1,2) f_{2}({\bm r}_{1}, {\bm v}_{1}, {\bm r}_{2}, {\bm v}_{2},t),
\end{equation}
where $ f_{1}({\bm r}_{1},{\bm v}_{1},t)$ and  $f_{2}({\bm r}_{1},{\bm v}_{1},{\bm r}_{2}, {\bm v}_{2}, t)$ are the one-particle and two-particle reduced distribution functions, respectively.  Attention will be restricted in this paper to spatially homogeneous states. A kinetic granular temperature $T(t)$ is defined from the mean square velocity of the particles according to
\begin{equation}
\label{2.10}
T(t) \equiv \frac{1}{n d} \int d{\bm v}_{1}\, m {\bm v}_{1}^{2} f_{1} ({\bm v}_{1},t),
\end{equation}
Above, $n \equiv N/V$ is the number of particles density, and it has been taken into account that the one-particle distribution function does not depend on the position for homogeneous states.

An equation for the time evolution of $T(t)$ can be derived by using
Eq. (\ref{2.9}),
\begin{equation}
\label{2.11}
\frac{\partial T(t)}{\partial t} = - \zeta(t) T(t),
\end{equation}
with the rate of change $\zeta (t)$ of the temperature due to the inelasticity of collisions given by
\begin{eqnarray}
\label{2.12}
\zeta(t) & = & - \frac{m \sigma^{d-1}}{nT(t)d} \int d{\bm v}_{1} \int d{\bm v}_{2}\, \int d \widehat{\bm \sigma}\, \theta ({\bm v}_{12} \cdot \widehat{\bm \sigma}) {\bm v}_{12} \cdot \widehat{\bm \sigma} \nonumber \\
&&\left[ \Delta^{2} + \alpha \Delta {\bm v}_{12} \cdot \widehat{\bm \sigma}- \frac{1- \alpha^{2}}{4} \left( {\bm v}_{12} \cdot \widehat{\bm \sigma} \right)^{2} \right] f_{2} ({\bm r}_{1},{\bm v}_{1},{\bm r}_{1}+ {\bm \sigma}, {\bm v}_{2},t).
\end{eqnarray}
For homogeneous systems $ f_{2} ({\bm r}_{1},{\bm v}_{1},{\bm r}_{1}+ {\bm \sigma}, {\bm v}_{2},t)$ does not depend on ${\bm r}_{1}$.

In place of the equilibrium state, the system has a stationary homogeneous state that has been discussed in \cite{BGMyB13}. In this paper, the time evolution of the temperature, eventually towards its steady value, under homogeneous conditions will be analyzed. On the appropriate time scale, the existence of an hydrodynamic-like description is expected. By definition, when the later holds, the time evolution of the temperature will obey a closed differential equation. The existence of this hydrodynamic regime will not be proven here, but it will  be checked {\em a posteriori} by comparing some of its predictions with numerical simulation results. A sufficient condition for homogeneous hydrodynamics is that the solution of the Liouville equation describing the time-dependent state of the system  be ``normal'', meaning that all its time dependence occurs through the granular temperature \cite{DyB11}. Then dimensional analysis requires that the probability density of the system $\rho_{H}(\Gamma, t)$ has the scaling form
\begin{equation}
\label{2.13}
\rho_{H}(\Gamma,t) = \left[ \ell v_{0}(t) \right]^{-Nd} \rho^{*}_{H} \left( \{ {\bm q}_{ij}, {\bm c}_{i}; i,j=1,\ldots, N \}, \Delta^{*} \right).
\end{equation}
The dimensionless function $\rho^{*}_{H}$ is invariant under space translations. From now on the subindex $H$ will be used to indicate that a quantity refers to the homogeneous hydrodynamic regime.  Velocities have been scaled relative to the thermal velocity and space coordinates relative to a quantity proportional to the mean free path,
\begin{equation}
\label{2.14}
{\bm q}_{i} \equiv \frac{{\bm r}_{i}}{\ell}, \quad
{\bm c}_{i} \equiv \frac{{\bm v}_{i}}{v_{0}(t)}\, ,
\end{equation}
with $\ell \equiv (n \sigma^{d-1})^{-1}$ and $v_{0}(t) \equiv \left( 2 T(t)/m \right)^{1/2}$. The dependence of $\rho^{*}$ on the dimensionless parameter
\begin{equation}
\label{2.15}
\Delta^{*} \equiv \frac{\Delta}{v_{0}(t)}
\end{equation}
has been indicated explicitly. In the hydrodynamic regime, the temperature obeys, by definition, a closed first order differential equation.  In \cite{BGMyB13} it was shown that there is only one steady temperature. Therefore, if the steady state is always reached in the homogeneous evolution, it follows that the temperature of the system increases or decreases depending on whether it is smaller or larger than its steady value, tending always monotonically to it.  Substitution of Eq. (\ref{2.13}) into the pseudo-Liouville equation (\ref{2.4}) gives
\begin{eqnarray}
\label{2.16}
\frac{\zeta^{*} (\alpha, \Delta^{*})}{2}  && \left\{ \sum_{i} \frac{\partial}{\partial {\bm c}_{i}}\, \cdot \left[ {\bm c}_{i}  \rho^{*}_{H} \left( \{ {\bm q}_{ij}, {\bm c}_{i} \}, \Delta^{*} \right) \right] + \Delta^{*} \frac{\partial}{\partial \Delta^{*}}\,  \rho^{*}_{H} \left( \{ {\bm q}_{ij}, {\bm c}_{i} \}, \Delta^{*} \right) \right\}
\nonumber \\
& & = \overline{L}_{+}^{*}(\Gamma^{*})  \rho^{*}_{H} \left( \{ {\bm q}_{ij}, {\bm c}_{i}\}, \Delta^{*} \right).
\end{eqnarray}
The dimensionless rate of change of the temperature is
\begin{eqnarray}
\label{2.17}
\zeta^{*}(\alpha, \Delta^{*}) & \equiv & \frac{\ell \zeta_{H}(t)}{v_{0}(t)}
\nonumber \\
&=& - \frac{2 \sigma^{* d-1}}{n^{*}d} \int d{\bm c}_{1} \int d{\bm c}_{2} \int d \widehat{\bm \sigma}\,  \theta \left( {\bm c}_{12} \cdot \widehat{\bm \sigma} \right) {\bm c}_{12} \cdot \widehat{\bm \sigma} \nonumber \\
&& \times \left[ \Delta^{*2} + \alpha \Delta^{*} {\bm c}_{12} \cdot \widehat{\bm \sigma}- \frac{1- \alpha^{2}}{4}\, \left( {\bm c}_{12} \cdot \widehat{\bm \sigma} \right)^{2} \right] f_{2,H}^{*}  \left({\bm q}_{1},{\bm c}_{1},{\bm q}_{1}+ {\bm \sigma}^{*},{\bm c}_{2}, \Delta^{*} \right),
\end{eqnarray}
with
\begin{equation}
\label{2.18}
\sigma^{*} = \frac{\sigma}{\ell}
\end{equation}
and
\begin{equation}
\label{2.19}
n^{*} =n \ell^{d}
\end{equation}
being the scaled diameter and number density, respectively. The dimensionless two-particle reduced distribution function is defined as
\begin{equation}
\label{2.20}
f^{*}_{2,H}({\bm q}_{1},{\bm c}_{1}, {\bm q}_{2},{\bm c}_{2}, \Delta^{*})= \left[ \ell v_{0}(t) \right]^{2d} f_{2,H} ({\bm r}_{1},{\bm v}_{1}, {\bm r}_{2},{\bm v}_{2},t).
\end{equation}
The pseudo-Liouville operator in Eq. (\ref{2.16}) is given by
\begin{equation}
\label{2.21}
\overline{L}^{*}_{+}(\Gamma^{*})= - \sum_{i} {\bm c}_{i} \cdot \frac{\partial}{\partial {\bm q}_{i}}+ \sum_{1 \leq i <j \leq N} \overline{T}^{*}_{+}(i,j),
\end{equation}
with the binary collision operator
\begin{eqnarray}
\label{2.22}
\overline{T}_{+}^{*}(i,j) & \equiv &  \sigma^{* d-1} \int d \widehat{\bm \sigma}\, \left[ \theta({\bm c}_{ij} \cdot \widehat{\bm \sigma} - 2 \Delta^{*}) ({\bm c}_{ij} \cdot \widehat{\bm \sigma}  - 2 \Delta^{*}) \delta ({\bm q}_{ij} -{\bm \sigma^{*}}) \alpha^{-2} b_{\bm \sigma}^{-1} (i,j) \right. \nonumber \\
& & - \left. \theta ( {\bm c}_{ij} \cdot \widehat{\bm \sigma} ){\bm c}_{ij} \cdot \widehat{\bm \sigma}  \delta ({\bm q}_{ij} + {\bm \sigma}^{*}) \right].
\end{eqnarray}
The operator $b_{\bm \sigma}^{-1}(i,j)$ is now understood to act on the velocities ${\bm c}_{i}$ and ${\bm c}_{j}$, changing them according to
\begin{equation}
\label{2.23}
b_{\bm \sigma}^{-1}(i,j) {\bm c}_{i} \equiv {\bm c}_{i} - \frac{1+ \alpha}{2 \alpha}\,  {\bm c}_{ij} \cdot \widehat{\bm \sigma}  \widehat{\bm \sigma} + \frac{\Delta^{*} \widehat{\bm \sigma}}{\alpha}\, ,
\end{equation}
\begin{equation}
\label{2.24}
b_{\bm \sigma}^{-1}(i,j) {\bm c}_{j} \equiv  {\bm c}_{j} + \frac{1+ \alpha}{2 \alpha}\, {\bm c}_{ij} \cdot \widehat{\bm \sigma}  \widehat{\bm \sigma} - \frac{\Delta^{*} \widehat{\bm \sigma}}{\alpha}\, .
\end{equation}
The above results in this section provide the natural representation to investigate homogeneous hydrodynamics in the model being considered here.

\section{Enskog approximation}
\label{s3}

The first equation of the hierarchy following by integration of Eq.\ (\ref{2.16}) is
\begin{eqnarray}
\label{3.1}
\frac{\zeta^{*} (\alpha, \Delta^{*})}{2}  && \left\{ \frac{\partial}{\partial {\bm c}_{1}} \cdot \left[ {\bm c}_{1} f_{1,H}^{*} ({\bm c}_{1},\Delta^{*} ) \right] + \Delta^{*} \frac{\partial }{\partial \Delta^{*}}\, f_{1,H}^{*} ({\bm c}_{1},\Delta^{*} ) \right\} \nonumber \\
&&= \int d{\bm q}_{2} \int d{\bm c}_{2}\, \overline{T}^{*}_{+}(1,2) f_{2,H}^{*}({\bm q}_{1},{\bm c}_{1},{\bm q}_{2},{\bm c}_{2}, \Delta^{*}),
\end{eqnarray}
where the dimensionless one-particle reduced distribution function is consistently defined as $f_{1,H}^{*}({\bm c}_{1}, \Delta^{*})= \left[ \ell v_{0}(t)\right]^{d} f_{1,H}({\bm v_{1}},t)$.

In the Enskog theory \cite{RydL77,Lu96,BGMyB13}, the precollisional two-body distribution function at contact is approximated as
\begin{eqnarray}
\label{3.2}
\lefteqn{\delta (r_{12}-\sigma) \theta (-{\bm v}_{12} \cdot \widehat{\bm \sigma} ) f_{2} ({\bm r}_{1},{\bm v}_{1},{\bm r}_{2},{\bm v}_{2},t)}
\nonumber \\
& & \approx  \delta (r_{12}- \sigma)  \theta (-{\bm v}_{12} \cdot \widehat{\bm \sigma} ) g_{E}[{\bm r}_{1},{\bm r}_{2} |n(t)] f_{1} ({\bm r}_{1}, {\bm v}_{1},t) f_{1} ({\bm r}_{2}, {\bm v}_{2},t).
\end{eqnarray}
Here, $g_{E}[{\bm r}_{1},{\bm r}_{2} |n(t)]$ is the equilibrium spatial pair correlation function evaluated with the non-equilibrium density field at time $t$. For the homogeneous states being considered here and in the dimensionless units introduced in the previous section, Eq.\ (\ref{3.2}) becomes
\begin{eqnarray}
\label{3.3}
\lefteqn{\delta (q_{12}-\sigma^{*}) \theta (-{\bm c}_{12} \cdot \widehat{\bm \sigma} ) f_{2,H}^{*} ({\bm q}_{1},{\bm c}_{1},{\bm q}_{2},{\bm c}_{2},\Delta^{*})} \nonumber \\
&& \approx \delta (q_{12}- \sigma^{*})g_{e}(\sigma,n)  \theta (-{\bm c}_{12} \cdot \widehat{\bm \sigma} ) )  n^{*2}\phi ({\bm c}_{1}, \Delta^{*}) \phi ({\bm c}_{2}, \Delta^{*}).
\end{eqnarray}
This expression involves the homogeneous equilibrium pair correlation function at contact, $g_{e}(\sigma,n) $, and the scaled velocity distribution $ \phi({\bm c}, \Delta^{*})$ defined by
\begin{equation}
\label{3.4}
f_{1,H}^{*}({\bm c},\Delta^{*}) = n^{*} \phi ({\bm c}, \Delta^{*}).
\end{equation}
Use of Eq. (\ref{3.3}) into Eq.\ (\ref{3.1}) leads to
\begin{equation}
\label{3.5}
\frac{\zeta^{*}_{B} (\alpha, \Delta^{*})}{2}   \left\{ \frac{\partial}{\partial {\bm c}} \cdot \left[ {\bm c} \phi ({\bm c},\Delta^{*} ) \right] + \Delta^{*} \frac{\partial }{\partial \Delta^{*}}\, \phi({\bm c},\Delta^{*} ) \right\}
= n^{*} J_{B}[{\bm c}|\phi],
\end{equation}
where $\zeta^{*}_{B}(\alpha, \Delta^{*}) $ is the Boltzmann limit of the dimensionless temperature rate,
\begin{eqnarray}
\label{3.6}
\zeta_{B}^{*}(\alpha, \Delta^{*}) & = & - \frac{2}{d} \int d{\bm c}_{1} \int d{\bm c}_{2} \int d \widehat{\bm \sigma}\ \theta \left( {\bm c}_{12} \cdot \widehat{\bm \sigma} \right) {\bm c}_{12} \cdot \widehat{\bm \sigma} \nonumber \\
&& \times \left[ \Delta^{*2} + \alpha \Delta^{*} {\bm c}_{12} \cdot \widehat{\bm \sigma}- \frac{1- \alpha^{2}}{4}\, \left( {\bm c}_{12} \cdot \widehat{\bm \sigma} \right)^{2} \right] \phi({\bm c}_{1}, \Delta^{*}) \phi ({\bm c}_{2}, \Delta^{*})
\end{eqnarray}
and $J_{B}[{\bm c}| \phi]$ is the (inelastic) Boltzmann collision term,
\begin{eqnarray}
\label{3.7}
J_{B}[{\bm c}_{1} | \phi] & = &\sigma^{*d-1} \int d{\bm c}_{2} \int d \widehat{\bm \sigma}\,  \left[ \theta({\bm c}_{12} \cdot \widehat{\bm \sigma} - 2 \Delta^{*}) ({\bm c}_{12} \cdot \widehat{\bm \sigma}  - 2 \Delta^{*}) \alpha^{-2} b_{\bm \sigma}^{-1} (1,2) \right. \nonumber \\
& & - \left. \theta ( {\bm c}_{12} \cdot \widehat{\bm \sigma} ){\bm c}_{12} \cdot \widehat{\bm \sigma}   \right] \phi ({\bm c}_{1}, \Delta^{*}) \phi ({\bm c}_{2}, \Delta^{*}).
\end{eqnarray}
It is worth to stress that Eq.\ (\ref{3.5}) is valid in the Enskog approximation, although it does not involves the pair correlation function. In this way, a closed differential equation for the one-particle velocity distribution has been obtained but, given its mathematical complexity, additional approximations are needed to be able to solve it by analytical methods. Then, the function $\phi({\bm c}, \Delta^{*})$ is expanded in  Sonine polynomials as \cite{RydL77}
\begin{equation}
\label{3.8}
\phi({\bm c}, \Delta^{*}) = \phi^{(0)} (c)  \sum_{j=0}^{\infty} a_{j}(\Delta^{*}) S^{(j)} (c^{2}),
\end{equation}
where
\begin{equation}
\label{3.9}
\phi^{(0)} (c) \equiv  \pi^{-d/2} e^{-c^{2}}
\end{equation}
and
\begin{equation}
\label{3.10}
S^{(j)} (c^{2}) \equiv  \sum_{r=0}^{j} \frac{\Gamma \left( j+d/2 \right)}{(j-r)! r! \Gamma \left( r+d/2 \right)}\, (-c^{2})^{r}.
\end{equation}
Normalization of the distribution and the scaling of the velocities with the thermal one imply that $a_{0}=1$ and $a_{1}=0$. Moreover, it is,
\begin{equation}
\label{3.11}
a_{2}(\Delta^{*})= \frac{4<c^{4}>}{d(d+2)}\, -1,
\end{equation}
with
\begin{equation}
\label{3.12}
<c^{4}> \equiv \int d{\bm c}\, c^{4} \phi ({\bm c},\Delta^{*}).
\end{equation}
In the following, the called first Sonine approximation,
\begin{equation}
\label{3.13}
\phi({\bm c}, \Delta^{*}) \approx \phi^{(0)} (c) \left[ 1+ a_{2} (\Delta^{*}) S^{(2)}(c^{2}) \right],
\end{equation}
will be used. Moreover, it will be assumed that $|a_{2} (\Delta)| \ll 1$, so that nonlinear terms in $a_{2}$ can be neglected, at least when computing low-order moments of $\phi({\bm c}, \Delta^{*})$. This assumption must be checked {\em a posteriori} and might imply a restriction on the intervals of values of $\alpha$ and $\Delta$ for which the obtained results are accurate.

When the approximation (\ref{3.13}) is employed in Eq.\ (\ref{3.6}) and the nonlinear in $a_{2}$ term is neglected, evaluation of the integrals gives
\begin{equation}
\label{3.14}
\zeta_{B}^{*}(\alpha, \Delta^{*}) \approx \frac{2^{3/2} \pi^{(d-1)/2}}{\Gamma \left( d/2 \right) d} \left[ \frac{1 - \alpha^{2}}{2} \left( 1+ \frac{3 a_{2}}{16} \right) - \alpha \left( \frac{\pi}{2} \right) ^{1/2} \Delta^{*} - \left( 1- \frac{a_{2}}{16} \right) \Delta^{*2} \right].
\end{equation}
In the steady state, the cooling rate must vanish, i.e.,
\begin{equation}
\label{3.15}
\zeta^{*}_{B}(\alpha, \Delta^{*}_{st})=0.
\end{equation}
In \cite{BGMyB13}, this equation was employed to calculate the steady temperature $T_{st}= m/2 \left( \Delta / \Delta^{*}_{st} \right)^{2}$. To determine $a_{2}(\alpha, \Delta^{*})$, Eq. (\ref{3.5}) must be used. Multiplication of that equation by $c^{4}$, integration over ${\bm c}$, and substitution of Eq. (\ref{3.14})  lead to a closed equation for $a_{2}$. The calculations are long and tedious, but they can be easily done by using any of the available software for symbolic calculation. If quadratic terms in $a_{2}$, as well as a term proportional to $a_{2} \partial a_{2}/ \partial \Delta^{*}$, are neglected, the result is
\begin{equation}
\label{3.17}
\frac{\partial a_{2}}{\partial \Delta^{*}} =  \left[ \frac{4}{\Delta^{*}}+ \frac{4 A_{1}+ B_{1}}{A_{0} \Delta^{*}} \right] a_{2} + \frac{4}{\Delta^{*}} + \frac{ B_{0}}{A_{0} \Delta^{*}} .
\end{equation}
The expressions of $A_{0}$, $A_{1}$, $B_{0}$, and $B_{1}$ are given in  the Appendix \ref{ap1}. The above differential equation can be numerically solved for a fixed value of $\alpha$ and a given initial condition $a_{2}(\alpha, \Delta^{*}_{0})=a_{2,0}$. It is important to realize the meaning of studying $a_{2}$ as a function of $\Delta^{*}$. As discussed above, $\Delta^{*}$ changes in time in a monotonic way, approaching its steady value. In the steady state the cooling rate vanishes by definition. The consequence is that the kinetic equation (\ref{3.5}) has a singularity at it, and this singularity translates to the equation for any velocity moment and, in particular, to the equation for $a_{2}$. Then, what has been done in the numerical resolution of the differential equation is to distinguish between trajectories starting with $ \Delta^{*}_{0} >\Delta^{*}_{st}$ and  with $ \Delta^{*}_{0} <\Delta^{*}_{st}$. In the former case the solution when $\Delta^{*}$ decreases was constructed, while in the later the solution for increasing $\Delta^{*}$ was considered. In both cases the calculations were stopped when approaching the singularity.  It is worth to point out that the singularity of Eq.  (\ref{3.17}) is not exactly located at $\Delta^{*}= \Delta^{*}_{st}$, but very close to it. The reason is that the contribution to the cooling rate proportional to $a_{2}$ in front of $\partial a_{2}/ \partial \Delta^{*}$ has been neglected.

\begin{figure}
\includegraphics[scale=0.4,angle=0]{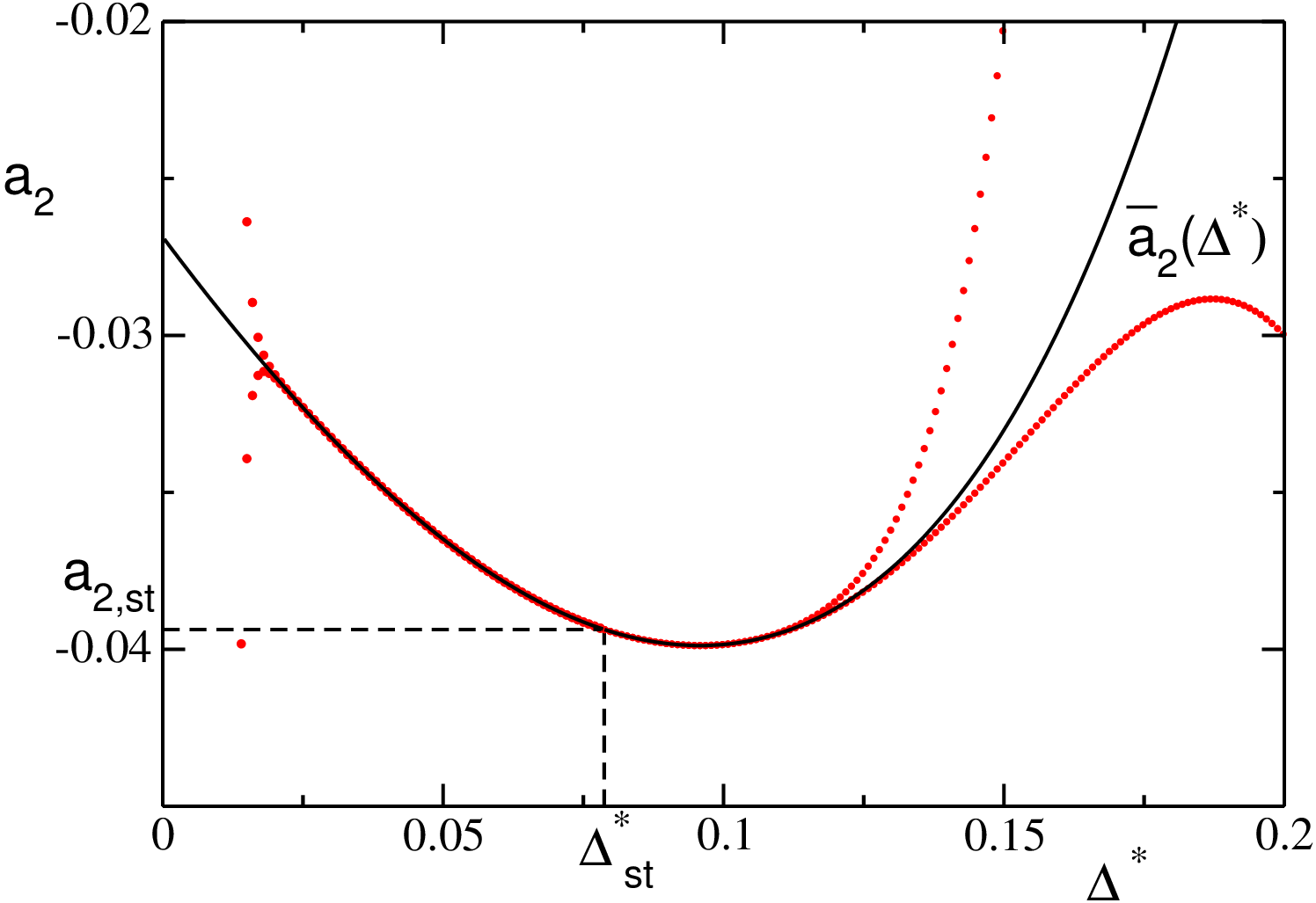}
\caption{(Color online) Sonine coefficient $a_{2}$ as a function of the dimensionless scaled velocity $\Delta^{*}$  for $\alpha=0.9$. The (red) symbols correspond to several numerical solutions of Eq. (\protect{\ref{3.17}}) obtained by using different initial conditions. The (black) solid line is the universal curve to which all the solutions converge.  This is precisely the function $\overline{a}_{2} (\Delta^{*})$, defining the normal one-particle distribution function  in the first Sonine approximation. Also indicated in the figure are the steady value of $a_{2}$, denoted by $a_{2,st}$, and the steady value of  $\Delta^{*}$, denoted by $\Delta_{st}^{*} $. \label{f1}}
\end{figure}

The numerical results obtained for $\alpha=0.9$ are shown in Fig.\ \ref{f1}. All the numerical trajectories converge very fast towards a universal curve $\overline{a}_{2} (\alpha,\Delta^{*})$, then forgetting the initial conditions from which they started. This is consistent with the assumption that the distribution function is normal. All the moments must depend on time only through $T(t)$, but they can not depend on the previous history or their initial values. Therefore, $\overline{a}_{2} (\alpha, \Delta^{*})$ is identified as the hydrodynamic expression of the second Sonine coefficient. The curve plotted in the figure has been obtained numerically, although it can be specified by saying that it is the only solution of the differential equation (\ref{3.17}) being finite for $\Delta^{*}$ going to zero.

\subsection*{Linear homogeneous hydrodynamics}
The hydrodynamic equation for the temperature in an homogeneous system is obtained by substituting the hydrodynamic expression, $\overline{a}_{2} (\alpha, \Delta^{*})$ of the Sonine coefficient into the expression of the cooling rate, Eq. (\ref{3.14}), and, afterwards, the latter into Eq. (\ref{2.11}). In this way, it is obtained:
\begin{equation}
\label{3.18}
\frac{\partial T(t)}{\partial s} = - g_{e} (\sigma,n) \overline{\zeta}^{*}_{B} (\alpha, \Delta^{*}) T(t),
\end{equation}
where $\overline{\zeta}^{*}_{B} (\alpha, \Delta^{*}) $ is  the hydrodynamic expression of the cooling rate, and a dimensionless time scale $s$ has been defined as
\begin{equation}
\label{3.19}
s= \int_{0}^{t} dt^{\prime}\, \frac{v_{0}(t^{\prime})}{\ell}.
\end{equation}
The time $s$ is proportional to the cumulative number of collisions per particle in the time interval between $0$ and $t$. The above equation for the temperature is a non-linear first order differential equation to be solved with some initial condition $T(0)$. For situations very close to the stationary value, $T_{st}$, the equation can be linearized about it to get
\begin{equation}
\label{3.20}
\frac{\partial}{\partial s}\, \frac{\delta T}{T_{st}} = -  \frac{\gamma \Delta^{*}_{st}}{2} \frac{\delta T}{T_{st}},
\end{equation}
with $\delta T(t) \equiv T(t)-T_{st}$ and
\begin{eqnarray}
\label{3.21}
\gamma (\alpha, \Delta^{*}_{st}) & = & -g_{e} (\sigma,n) \left( \frac{\partial \overline{\zeta}^{*}_{B}(\alpha, \Delta^{*})}{\partial \Delta^{*}} \right)_{\Delta^{*}= \Delta^{*}_{st}} \nonumber \\
&=& \frac{2^{3/2} \pi^{\frac{d-1}{2}}g_{e} (\sigma,n)}{\Gamma \left( d/2 \right)d} \left[ \left( \frac{\pi}{2} \right)^{1/2} \alpha + 2\left( 1- \frac{a_{2,st}}{16} \right) \Delta^{*}_{st} \right].
\end{eqnarray}
Equation (\ref{3.20}) leads to identify the hydrodynamic eigenmode
\begin{equation}
\label{3.22}
\lambda = \frac{\gamma \Delta^{*}_{st}}{2},
\end{equation}
that is always positive, indicating the linear stability of the hydrodynamic equation. Moreover, since $\Delta_{st}^{*}$ formally vanishes in the elastic limit, it is
\begin{equation}
\label{3.23}
\lim_{\alpha \rightarrow 1} \lambda =0.
\end{equation}
This reflects that, for given $\Delta$, the decay of an hydrodynamic perturbation slows down as the system becomes more elastic. If the Gaussian approximation is used in Eq. (\ref{3.21}) by putting $a_{2,st}=0$, a very accurate expression is obtained, 
\begin{equation}
\label{3.24}
\gamma (\alpha, \Delta^{*}_{st}) \simeq \frac{2 \pi^{\frac{d}{2}}g_{e} (\sigma,n)}{\Gamma \left( d/2 \right)d} \sqrt{ \alpha^{2} + \frac{4(1-\alpha^{2})}{\pi}},
\end{equation}
consistently with the results reported above for the hydrodynamic decay of the temperature to its stationary value.

\section{Direct Monte Carlo simulations}
\label{s4}
The direct simulation Monte Carlo (DSMC) method was devised to mimic the dynamics of a low density gas described by the Boltzmann equation \cite{Bi94,Ga00}. It provides a very efficient tool to generate numerical solutions of the Boltzmann equation, and also to get information on the fluctuations and correlations present in the gas. One of the advantages of the method is that it allows to explote in a direct way the symmetries of the state under study. For homogeneous systems, if the attention is restricted to the one-particle distribution function, there is no need to consider the spatial coordinates of the particles. Consistently, no boundary conditions must be introduced. Of course, this does not hold if the aim of the study would include spatial correlations. A point to stress is that for homogeneous systems, the only difference between the Boltzmann and the Enskog equation is the presence of the equilibrium pair correlation function at contact as a constant prefactor in the collision term of the latter. This constant can always be eliminated by redefining the time scale and, as a consequence, any solution of the homogeneous Boltzmann equation is directly related to a solution of the homogeneous  Enskog equation.

We have applied  the DSMC method to a two dimensional system, since this is the dimension for which the model being considered is expected to be more relevant, in the sense of modeling the horizontal dynamics of a vibrated gas of inelastic hard spheres,  confined to a quasi-two dimensional geometry \cite{BRyS13,BGMyB13}.  The number of particles employed in the simulations  is $N=1000$, and the reported results have been averaged over 5000  trajectories.

\begin{figure}
\includegraphics[scale=0.4,angle=0]{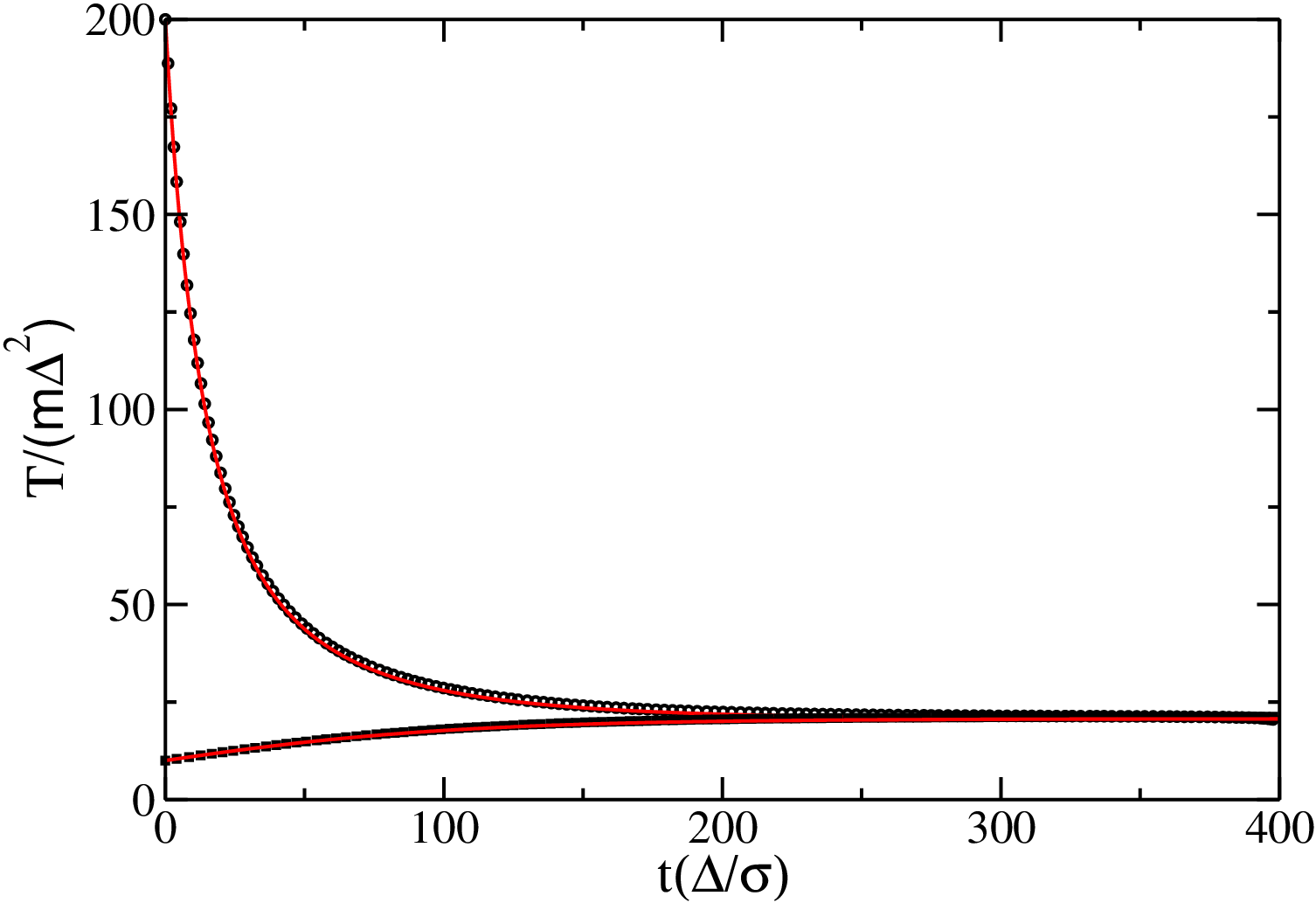}
\caption{(Color online)  Time evolution of the temperature in a confined system with a coefficient of normal restitution $\alpha= 0.8.$ Both the temperature and the time are measured in the dimensionless units indicated in the labels of the axis. The (black) symbols are DSMC method results obtained with two different initial temperatures and the (red) solid line is the theoretical prediction derived in the text, i.e. the solution of Eq. (\protect{\ref{2.11}}) using Eq.  (\protect{\ref{3.14}}). \label{f2}}
\end{figure}

In Fig. \ref{f2}, the time evolution of the dimensionless temperature $T(t)/m \Delta^{2}= 1/2 \Delta^{*2}$ is shown for a system with $\alpha=0.8$. The initial velocity distribution is Gaussian and results for two different initial granular temperatures, $T(0)= 200 m \Delta^{2}$ and $T(0) = 10 m \Delta^{2}$, are plotted. One of them is above the steady temperature and the other one below it. Time is measured in units of $\sigma / \Delta$. In the units used, the dynamics of the particles does not depend on $\Delta$ \cite{BGMyB13} and, consistently, the same happens with the equation for the evolution of the temperature derived in this paper. The symbols in the figure are simulation results and the solid line is the theoretical prediction, i.e. the solution of Eq. (\ref{2.11}) with the temperature rate given by Eq. (\ref{3.14}). A very good agreement between theory and simulation is observed. The absence of a significant initial transient in the simulation behaviour before reaching the hydrodynamic curve, is because the velocity distribution function in the hydrodynamic regime is close to the Gaussian used as initial condition in the simulations. Similar results have been obtained for other values of the coefficient of normal restitution.

\begin{figure}
\includegraphics[scale=0.4,angle=0]{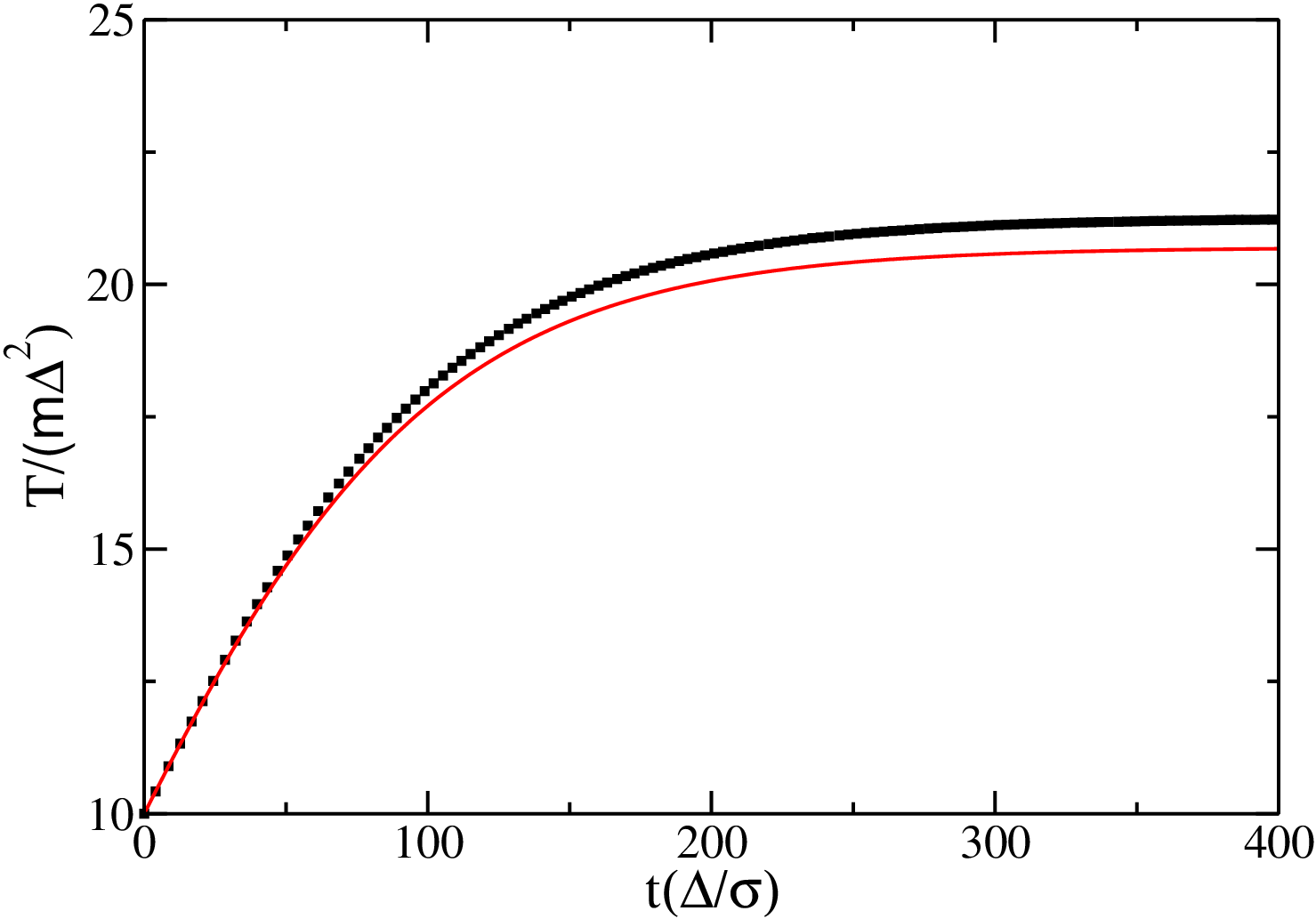}
\caption{(Color online)  Time evolution of the temperature in a confined system with a coefficient of normal restitution $\alpha= 0.8$ and an initial temperature $T(0)= 10 m \Delta^{2}$.  Both the temperature and the time are measured in the dimensionless units indicated in the labels of the axis. The (black) symbols are DSMC method results  and the (red) solid line is the theoretical prediction derived in the text, Eq. (\protect{\ref{2.11}}), putting $a_{2}=0$. \label{f3}}
\end{figure}

On the scale used in Fig. \ref{f2}, the influence of the term proportional to $a_{2}$ in Eq. (\ref{3.14}) can not be made out. To show that this term can be clearly identified in the simulation results, in Fig.\ \ref{f3} the time evolution of the temperature in one of the cases reported in Fig. \ref{f2}  is again plotted, now on a  larger scale. The solid line has been obtained by using  Eq. (\ref{3.14})  with $a_{2}=0$. The discrepancy with the simulation data is clearly identified. In ref. \cite{BGMyB13},  it was shown that the theoretical prediction for the steady value of $a_{2}$ agree fairly well with the simulation values. Therefore, such comparison will be not repeated here

\begin{figure}
\includegraphics[scale=0.4,angle=0]{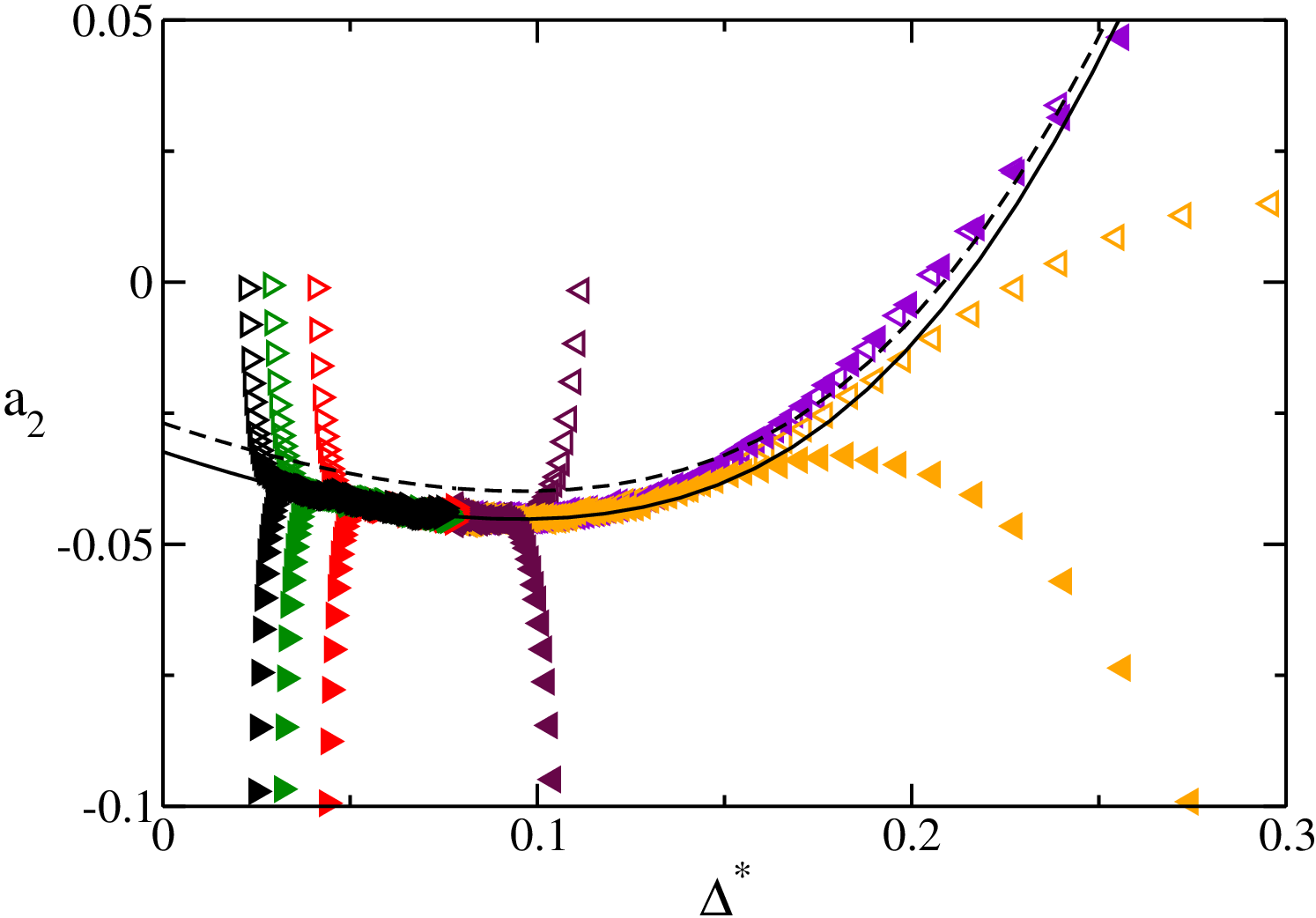}
\caption{(Color online) Sonine coefficient $a_{2}$ as a function of the dimensionless velocity $\Delta^{*}$ for a system with $\alpha=0.9$ . The symbols are simulation results obtained by the DSMC method. Two different kinds of initial conditions have been used:  Gaussian velocity distributions (empty symbols) and square distributions (filled symbols). The dashed line is the theoretical prediction obtained by solving Eq.\ (\ref{3.17})  and considering the hydrodynamic regime, as discussed in the main text.  Finally, the solid line is just the translation of the latter so that the position, $\Delta^{*}$,  of the steady value and the value itself, $\overline{a}_{2,st}$,  coincide with the simulation results.  \label{f4}}
\end{figure}

The time evolution of the Sonine coefficient $a_{2}(\alpha)$, and the existence of a homogeneous hydrodynamic regime have also been studied in the simulations. The coefficient has been evaluated using the exact relationship given in Eq. (\ref{3.11}), i.e. what has been actually measured is the fourth moment of the velocity distribution. Two kind of initial velocity distributions have been used: Gaussians and square distributions, both with zero mean average. Of course, the first ones correspond to a vanishing initial value of $a_{2}$. Note that in some cases, an initial interval of values of $\Delta^{*}$ lies outside the plotted interval. The results obtained for $\alpha=0.9$ are shown in Fig. \ref{f4}. In the simulations, the variation of $\Delta^{*}$ is due to the (monotonic) change in time of the temperature, approaching its steady value. In all the cases, the Sonine coefficient tends to a universal, normal curve, afterwards approaching its steady value. The theoretical prediction $\overline{a}_{2} (\Delta^{*})$, as given in Fig.\  \ref{f1} is included (dashed line). Moreover, also plotted (solid line) is the curve obtained by translating the theoretical prediction, so that the steady simulation value of the Sonine coefficient coincides with the theoretical prediction for it.  The discrepancy of the dashed line is probably due to the linear in $a_{2}$ approximation used when deriving Eq. (\ref{3.17}),  as discussed in the previous Section. A very good agreement is  obtained. Similar results have been obtained for $\alpha=0.8$. These results confirm the existence of homogeneous hydrodynamics for the collisional model, and also the consistency of assuming a normal distribution function to describe it.

\section{Final remarks}
\label{s5}

In this paper, the macroscopic equation for the homogeneous evolution of the collisional model recently proposed to describe the two-dimensional dynamics of confined granular gases \cite{BRyS13,BGMyB13}, has been investigated. It has been shown that there is a hydrodynamic regime in which the evolution of the temperature is accurately described by a closed first order differential equation. In the context of non-equilibrium statistical mechanics, hydrodynamics is characterized by a {\em normal} distribution function of the system, having the peculiarity that all its time dependence occurs through the (granular) temperature. Thus dimensional analysis implies some scaling of the  distribution function. The precise form of the scaling is a consequence of the formulation of the model. The latter involves a characteristic velocity,  $\Delta$, describing the way in which the energy is transferred from the vertical degree of freedom to the horizontal ones through the collisions between particles. That means that  two independent dimensionless quantities can be made by scaling with the thermal velocity, one from the  particles velocity, ${\bm v}$, and  another from the characteristic speed, $\Delta$. In general, the distribution function of the normal state can depend on both, and this must be taken into account when deriving the form of the hydrodynamic equations. It is important to stress that this is not a peculiarity of the model discussed here. A similar behavior has been found in granular systems driven by an external stochastic thermostat \cite{GMyT12,GMyT13}. With more generality, the same can be expected when studying hydrodynamics around a steady state of a molecular fluid \cite{Lu06}. The reason is that the characterization of the steady state necessarily involves some gradient of a hydrodynamic field and, therefore, an additional magnitude to be scaled out with some function of the temperature. 

Although the dependence of the normal distribution function on the scaled characteristic velocity, $\Delta^{*}$, plays no relevant quantitative role in the macroscopic evolution of the temperature or in its steady value,  it is fundamental to develop a formally consistent theory. Only taken that dependence into account, the kinetic equation admits a normal solution.  In any case, this new dependence will clearly show up also when studying the hydrodynamic fluxes in the system and it will affect the form of the transport coefficients. It might happen that it be quite relevant in this context, and gave quantitatively important corrections to the transport coefficients, affecting even the stability analysis.

Another relevant issue is that the time-dependent normal distribution function describing the homogeneous hydrodynamics can not be inferred from the distribution of the steady state, since in the latter the temperature is fixed by some condition, namely the vanishing of the rate of change of the temperature. When deriving hydrodynamic equations by means of an expansion in the gradients of the hydrodynamic fields, e.g. by an extension of the Chapman-Enskog method \cite{DyB02}, the zeroth order distribution must describe an state with arbitrary uniform deviations of all the hydrodynamic fields. In other worlds, it is the hydrodynamic time-dependent homogeneous state discussed in this paper the one to be used.

\section{Acknowledgements}

This research was supported by the Ministerio de Educaci\'{o}n y Ciencia (Spain) through Grant No. FIS2011-24460 (partially financed by FEDER funds).

\appendix*

\section{Equation for the second Sonine coefficient}
\label{ap1}
The coefficient $a_2(\alpha, \Delta^{*})$ appearing  in Eq.\  (\ref{3.13}) obeys, in the linear approximation, Eq.\ (\ref{3.15}), where
\begin{equation}
\label{ap1.1}
A_{0} (\alpha, \Delta^{*})= (d+2) \left[ \frac{1- \alpha^{2}}{2} - \left( \frac{\pi}{2} \right)^{1/2}  \alpha \Delta^{*} - \Delta^{*2} \right],
\end{equation}
\begin{equation}
\label{ap1.2}
A_{1} (\alpha, \Delta^{*}) = \frac{(d+2)}{16} \left[ \frac{3(1- \alpha^{2})}{2}\, + \Delta^{*2} \right],
\end{equation}
\begin{eqnarray}
\label{ap1.3}
B_{0}(\alpha, \Delta^{*}) &=& (2 \pi )^{1/2} \left(1+2d+3 \alpha^{2} +4 \Delta^{*2}\right) \alpha \Delta^{*}-3+4 \Delta^{*4} + \alpha^{2}+ 2 \alpha^{4}  \nonumber  \\
&& -2d \left( 1-\alpha^{2}-2 \Delta^{*2} \right)+ 2 \Delta^{*2} \left( 1+6 \alpha^{2} \right),
\end{eqnarray}
\begin{eqnarray}
\label{ap1.4}
B_{1}(\alpha,\Delta^{*})& = & \left( \frac{\pi}{2} \right)^{1/2} \left[ 2 -2d(1-\alpha)+7 \alpha+3 \alpha^{3} \right] \Delta^{*}- \frac{1}{16} \left\{ 85 + 4 \Delta^{*4}-18 (3+2 \alpha^{2}) \Delta^{*2} \right.  \nonumber \\
&& \left. - \left( 32 +87 \alpha+ 30 \alpha^{3} \right) \alpha -2d \left[ 6 \Delta^{*2}-(1+\alpha) (31-15 \alpha) \right] \right\}.
\end{eqnarray}


\begin{thebibliography}{99}

\bibitem{Ha83} P.K. Haff, J. Fluid Mech. {\bf 134}, 401 (1983).

\bibitem{Go03} I. Goldhirsch, Annu. Rev. Fluid Mech. {\bf 35}, 267 (2003).

\bibitem{ByP04} N. V. Brilliantov and T. P\"{o}schel, {\em Kinetic Theory of Granular Gases} (Oxford University Press, Oxford, 2004).

\bibitem{Du00} J.W. Dufty, J. Phys.: Condens. Matter {\bf 12}, A47 (2000).

\bibitem{DByB08} J.W. Dufty, A. Baskaran, and J.J. Brey, Phys. Rev. E {\bf 77}, 031310 (2008).

\bibitem{BDKyS98} J.J. Brey, J.W. Dufty, C.S. Kim, and A. Santos, Phys. Rev. E {\bf 58}, 4638 (1998).

\bibitem{ByC01} J.J. Brey and D. Cubero, in {\em Granular Gases}, edited by T. P\"{o}schel and S. Luding (Springer-Verlag, Berlin 2001).

\bibitem{OyU05} J.S. Olafsen and J.S. Urbach, Phys. Rev. Lett. {\bf 95}, 098002 (2005).

\bibitem{RPGRSCyM11} N. Rivas, S. Ponce, B. Gallet, D. Risso, R. Soto, P. Cordero, and N. M\'{u}jica, Phys. Rev. Lett.\ {\bf 106}, 088001 (2011).

\bibitem{vNyE98} T.P.C. van Noije and M.H. Ernst, Granular Matter {\bf 1}, 57 (1998).

\bibitem{PLMyV99} A. Puglisi, V. Loreto, U. Marini Bettolo Marconi, and A. Vulpiani, Phys. Rev. E {\bf 59}, 5582 (1999).

\bibitem{PTvNyE01} I. Pagonabarraga, E. Trizac, T.P.C. van Noije, and M.H. Ernst, Phys. Rev. E {\bf 65},  011303 (2001).

\bibitem{BRyS13} R. Brito, D. Risso, and R. Soto, Phys. Rev. E {\bf 87}, 022209 (2013).

\bibitem{BGMyB13} J.J. Brey, M.I. Garc\'{\i}a de Soria, P. Maynar, and V. Buz\'{o}n, Phys. Rev. E {\bf 88}, 062205 (2013).

\bibitem{GMyT12} M.I. Garc\'{\i}a de Soria, P. Maynar, and E. Trizac, Phys. Rev. E {\bf 85}, 051301 (2012).

\bibitem{GMyT13} M.I. Garc\'{\i}a de Soria, P. Maynar, and E. Trizac, Phys. Rev. E {\bf 87}, 022201 (2013).

\bibitem{Lu06} J. F. Lutsko, Phys. Rev. E {\bf 73}, 021302 (2006).

\bibitem{DyB11} J.W. Dufty and J.J. Brey, Math. Model. Nat. Phenom. {\bf 6}, 19 (2011).

\bibitem{RydL77} P. R\'{e}sibois and M. de Leener, {\em Classical Kinetic  Theory of Fluids} (John Wiley \& Sons, New York, 1977).


\bibitem{Lu96} J.F. Lutsko, Phys. Rev. Lett. {\bf 77}, 2225 (1996).

\bibitem{Bi94} G. Bird, {\em Molecular Gas Dynamics and the Direct Simulation of Gas Flows}  (Clarendom, Oxford, 1994).

\bibitem{Ga00} A. Garc\'{\i}a, {\em Numerical Methods for Physics} (Prentice Hall, Englewood Cliffs, NJ, 2000).

\bibitem{DyB02} J.W. Dufty and J.J. Brey, J. Stat. Mech. {\bf 109}, 433 (2002).


\end{thebibliography}
\end{document}